\documentstyle[12pt,epsf]{article}
\begin{document}
\baselineskip=0.7cm
\begin{titlepage}
\begin{flushright}
WM-97-103\\
September 14, 1997
\end{flushright}
\bigskip

\begin{center}
{\Large \bf Quark-Antiquark Bound States within a Dyson-Schwinger 
Bethe-Salpeter Formalism}
\bigskip
\bigskip

\c{C}etin \c{S}avkl{\i}$^{a,}$\footnote{E-mail: csavkli@physics.wm.edu}, 
Frank Tabakin$^{b,}$\footnote{E-mail: frankt@tabakin.phyast.pitt.edu}
\medskip

{\it $^a$Department of Physics, College of William and Mary, Williamsburg, 
Virginia 23185,\\
$^b$Department of Physics \& Astronomy, University of Pittsburgh, 
Pittsburgh, Pennsylvania  15260}
\medskip 
\end{center}
\bigskip

\bigskip

\begin{abstract}
Pion and kaon observables are calculated using a Dyson-Schwinger 
Bethe-Salpeter formalism. It is shown that an infrared finite gluon 
propagator can lead to quark confinement via generation of complex mass poles 
in quark propagators. Observables, including electromagnetic form factors, 
are calculated entirely in Euclidean metric for spacelike values of bound 
state momentum and final results are extrapolated to the physical region. 

\end{abstract}
\vfill



\end{titlepage}

\section{Introduction}
\label{sec:introduction}
Description of simple hadrons in terms of quark-gluon degrees of freedom 
has long been an active area in physics. With the advent of TJNAF, which 
will be operating at intermediate energies and therefore probing the structure 
of hadrons, there is new motivation and need for a simple theoretical 
description of quark interactions. In this context, the Dyson-Schwinger 
Bethe-Salpeter(DSBS) equation formalism has gained popularity in recent 
years.\footnote{See Ref.~\cite{ROBERTS1} for an extensive review.} The DSBS 
formalism serves to bridge the 
gap between nonrelativistic quark models and more rigorous approaches, such as 
lattice gauge theory. 

The main features of QCD  can be summarized as chiral symmetry breaking, 
confinement and asymptotic freedom. It is possible to address all of these 
features within the DSBS formalism. In this formalism, the input is an 
effective gluon propagator which is assumed to represent the
interactions between quarks at all momentum transfers. The choice of a vector 
interaction between quarks is motivated only by the desire to make a 
connection with QCD degrees of freedom. In fact, whether a scalar or a vector 
interaction should be used between quarks is a topic of debate not 
addressed in this paper. 

While various applications of the DSBS formalism to pseudoscalar and vector 
mesons have produced promising results, there are still some questions to be 
investigated.  In this paper,  we address three issues. These are: {\bf a)} 
Can an infrared gluon propagator lead to confined quarks? {\bf b)} The 
question of using Euclidean metric and extrapolation. {\bf c)} Dressing
of the quark-gluon vertex in the DSBS equations while maintaining the chiral 
limit ? 

The organization of this paper is as follows: In section \ref{section2}, the 
model is introduced and the realization of the chiral limit, within the 
dressing scheme used in this paper, is discussed. In section 
\ref{section3}, the quark propagator functions obtained by solving the 
Dyson-Schwinger equation are presented and the quark propagator is shown to be
free of real timelike poles, indicating that quarks can not be free, which is
an implication and requirement of confinement. In section \ref{section4}, 
meson and kaon observables, which are calculated using a Euclidean 
metric(rather than a Wick rotation that only effects the internal momenta),  
are presented. Finally, results are summarized and our conclusions are 
presented in section \ref{section5}.  

\section{THE MODEL}
\label{section2}

\begin{figure}
\begin{center}
\mbox{
   \epsfxsize=3.2in
   \epsfysize=0.6in 
\epsffile{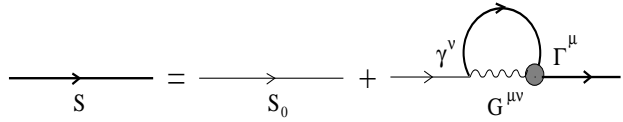}
}
\end{center}
\caption{Quark Dyson-Schwinger equation is shown. Only one vertex is dressed 
to prevent double counting.}
\label{ds.fig}
\end{figure}

\begin{figure}
\begin{center}
\mbox{
   \epsfxsize=3.2in
   \epsfysize=0.8in 
\epsffile{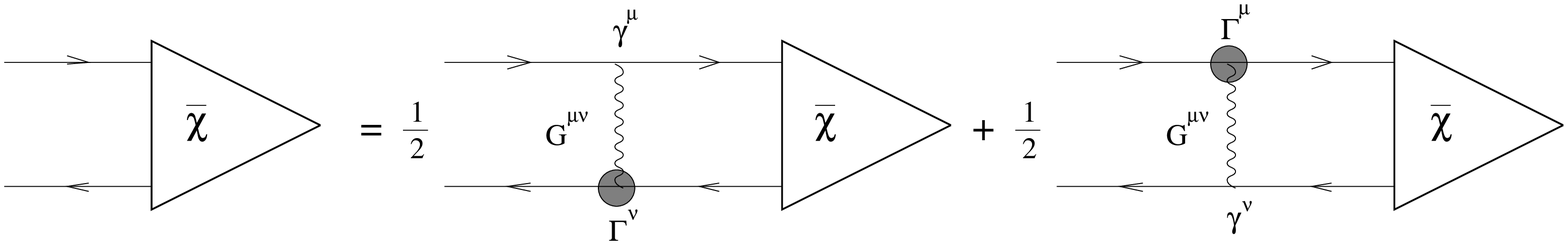}
}
\end{center}
\caption{The Bethe-Salpeter equation is shown. Only one vertex is dressed 
each time. This is necessary to preserve the chiral limit, which follows from 
the similarity of the BS and the DS equations.}
\label{bs.fig}
\end{figure}

The mechanism of chiral symmetry breaking and recovery of massless pseudoscalar
bound states(pion, kaon) in the limit of massless fermions(quarks) was 
originally discovered in the papers of Nambu-Jona-Lasinio(NJL)~\cite{NJL1}. 
Nambu-Jona-Lasinio's model originally described 
relativistic nucleon interactions through 
local, four-nucleon couplings. It is the same philosophy 
that is followed in the DSBS calculations, except that now nucleons are 
replaced with quarks and the contact interaction is replaced by an effective 
gluon exchange between quarks. The Dyson-Schwinger(DS) and the 
Bethe-Salpeter(BS) equations employed in this work are shown in 
Figures~\ref{ds.fig} and~\ref{bs.fig}. The DS equation describes the 
propagation of quarks in the presence of gluons. The BS equation 
 describes the quark-antiquark bound state, in which the DS quark 
propagator is used.
  
In Fig.~\ref{ds.fig}, the thin lines represent the current quark propagators 
for each flavor f:
\begin{equation}
i\,S^0_f(p)=\Bigl[\frac{i}{\ \! / \! \! \! p -m_f^0}\Bigr],
\end{equation} 
while the thick lines correspond to the dressed quark propagators: 
\begin{equation}
i\,S_f(p)=\Bigl[\frac{i}{A_f(p)\ \! / \! \! \! p -B_f(p)}\Bigr].
\end{equation} 
Here $A_f(p)$ is a dimensionless normalization factor and $B_f(p)$ has the 
units of mass(MeV).

The problem of how to systematically dress DS and BS equations has recently 
been addressed\cite{MUNCZEK,ROBERTS6}. Dressing of all vertices consistently 
is motivated by the desire to make a closer connection with QCD. The 
structure of the Dyson-Schwinger equation, when combined with the chiral limit
requirement, strictly restricts the choice of kernel for the Bethe-Salpeter 
equation. The chiral limit in NJL type models~\footnote{See 
Refs.~\cite{KLEVANSKY,KVW} for an extensive review of the NJL type models, and 
Refs.~\cite{SHAKIN} for an extended version of it.} is obtained due to the 
similarity of  the Dyson-Schwinger and Bethe-Salpeter equations in the limit 
of massless current quarks. In this limit,  the quark mass function $B(p)$ and
the Bethe-Salpeter wavefunction $\Phi(p)$ for pseudoscalar massless bound 
states satisfy the same equation(to be shown below). Therefore, for any given 
set of parameters of the model gluon propagator $G^{\mu\nu}$, the solution of 
the Dyson-Schwinger equation automatically implies a massless pseudoscalar 
bound state solution for the Bethe-Salpeter equation. In other words, in the 
chiral limit the quark Dyson-Schwinger equation produces the appropriate mass 
function $B(p)$ such that the BS equation produces a massless pseudoscalar 
bound state. 

In order to prevent double counting, only one of the vertices in the 
Dyson-Schwinger equation(Fig.~\ref{ds.fig}) is dressed as indicated by the 
solid circle. Therefore,  to preserve the similarity between 
the BS and the DS equations in the chiral limit, we dress only one of the 
quark-gluon vertices in constructing the BS equation. In order to keep the 
Bethe-Salpeter equation symmetric(to treat quarks equally),  the kernel is 
divided into two pieces, where in each piece an alternate vertex is dressed, 
and contribution of those terms is averaged(See Fig.~\ref{bs.fig}). While the 
dressing of only one of the quark-gluon vertices in the BS kernel does not 
represent a complete dressing, since cases where both quark-gluon vertices are
simultaneously dressed are excluded, with the proper choice of vertex 
$\Gamma^{\mu}$, one has a {\em subset} of all diagrams that produce the 
correct chiral limit.
  
Having stated the general structure of the DS and  BS equations used in this
calculation, we now discuss the chiral limit, and the choice of the gluon 
propagator $G^{\mu\nu}(q)$.
    
In terms of quark and gluon propagators, the DS equation is written as 
\begin{eqnarray}
&&S_f(p)=S_{0f}(p)+i\frac{4}{3}S_f(p)\bigl[\int\,\frac{d^4q}{(2\pi)^4}\,
G^{\mu\nu}(p-q)\,\Gamma_{\mu}(p,q)S_f(q)\gamma_{\nu}\bigr]S_{0f}(p).\label{sd1}
\end{eqnarray}
Similarly, the Bethe-Salpeter equation~\cite{BETHESALPETER} determining the BS
 vertex function(a truncated wavefunction, see later) $\chi_{P}(k)$ is given 
by 
\begin{eqnarray}
\overline{\chi}_{P}(k)&=&i\frac{4}{3}\int\,\frac{d^4q}{(2\pi)^4}\,G^{\mu\nu}
(q-k)\,
\frac{1}{2}\Biggl[\Gamma_{\nu}(-k_-,-q_-)\,S_d(-q_-)\,\overline{\chi}_P(q)\,
S_u(q_+)\,
\gamma_{\mu}\nonumber\\
&&+\gamma_{\nu}\,S_d(-q_-)\,\overline{\chi}_P(q)\,S_u(q_+)\,\Gamma_{\mu}
(q_+,k_+)
\Biggr]\label{bseq1},
\end{eqnarray}
where the 4-vector $q_+=P\eta_1+q$, $q_-=P\eta_2-q$, $\eta_1+\eta_2=1$, and 
$P$ is the bound state 4-momentum. The BS vertex function $\chi_{P}(k)$  and 
its conjugate $\overline{\chi}_{P}(k)$ are related~\cite{MANDELSTAM} by  
\begin{equation}
\overline{\chi}_P(ik_0,\vec{k})=\gamma_0\chi^*_P(-ik_0,\vec{k})\gamma_0.
\end{equation}
The normalization condition of the BS vertex function is derivable from the 
BS equation itself.~\cite{ITZYKSON} If the interaction does not depend on the 
total bound state momentum, the normalization condition reduces to
\begin{eqnarray}
2P^{\mu}&=&iN_c\int\,\frac{d^4q}{(2\pi)^4}\,{\rm tr_{D}}\Biggl[
\overline{\chi}_P(q)\,\frac{\partial}{\partial P_{\mu}}S(q_+)\,\chi(q)\,
S(-q_-)\nonumber\\
&&+\overline{\chi}_P(q)\,S(q_+)\,\chi(q)\,\frac{\partial}{\partial P_{\mu}}
S(-q_-)\Biggr]
\label{bsnorm}
\end{eqnarray}
As it is well known, the dressing of electromagnetic vertices, such as the 
photon-quark vertex, is constrained by the Ward-Takahashi identity,
\begin{equation}
q_{\mu}\Gamma^{\mu}(p',p)=S^{-1}(p')-S^{-1}(p),
\label{wtid}
\end{equation}
 which guarantees the conservation of electromagnetic current at the vertex.
Similarly, due to color current conservation, the dressed quark-gluon 
interaction vertex, $\Gamma^{\mu}(p,q)$, 
satisfies the Slavnov-Taylor identity~\cite{MARCIANO}
\begin{eqnarray}
&&q_{\mu}\Gamma^{\mu}(p',p)[1+b(q^2)]=[1-B(q,p)]S^{-1}(p')-S^{-1}(p)[1-B(q,p)],
\label{sltay}
\end{eqnarray}
where functions $b(q^2)$ and $B(q,p)$ are related to 
``ghost fields.'' Since the QCD ghost field contributions to Eq.~\ref{sltay} 
are not well 
understood we neglect their contributions.
When ghost fields are neglected, $b(q^2)$ and $B(q,p)=0$; the Slavnov-Taylor 
identity then reduces to the Ward-Takahashi identity Eq.~\ref{wtid}. The 
minimal 
vertex that satisfies this identity in this limit has been given by
Ball-Chiu~\cite{BALLCHIU} as:
\begin{eqnarray}
&&\Gamma^{\mu}_{BC}(p',p)=\frac{A(p')+A(p)}{2}\gamma^{\mu}+\frac{(p'+p)^{\mu}}
{p'^2-p^2}\Bigl[\frac{A(p')-A(p)}{2}\,(\ \! / \! \! \! p'+\ \! / \! \! \! p)
+B(p')-B(p)\Bigr].\label{BallChiu}
\end{eqnarray} 
It is clear that one can add any term to this vertex  that satisfies 
\begin{equation}
q_{\mu}\Gamma^{\mu}(p',p)=0.
\end{equation} 
Curtis-Pennington~\cite{CURTPENN3} have proposed such an additional vertex 
term. Here, 
for simplicity, we consider only the Ball-Chiu dressing.

\subsection{The Chiral Limit}
To determine what type of quark-gluon vertex is allowed within the 
approximation scheme employed here, let us analyze the chiral limit
of the DS and the BS equations. In the chiral limit the current quark masses
$m_0$ and the bound state momentum $P$ vanishes. In this limit, taking the 
spinor trace of the Dyson-Schwinger 
equation~\ref{sd1}, one obtains   
\begin{eqnarray}
B(p)&=& \frac{i}{3} \int  \frac{d^4k}{(2\pi)^4}\, G_{\mu \nu}(p-k)\,
\frac{B(k)}{A^2(k)k^2-B^2(k)}\, {\rm tr}\Bigl[\Gamma^{\mu}(p,k)\,
(1+\frac{A(k)}{B(k)}\ \! / \! \! 
\! k)\,\gamma^{\nu}\Bigr].
\label{bfunct}
\end{eqnarray}
For a pseudoscalar bound state, in the chiral limit the BS vertex function is 
given by  
\begin{equation}
\chi_P(k)\equiv i\gamma_5\Phi_P(k),\nonumber
\end{equation}
where $\Phi_P(k)$ is a scalar function of the relative momentum $k$, and bound
state momentum $P=0$.
With this definition, the BS equation for the bound state wavefunction 
$\Phi_P(k)$ is obtained as
\begin{eqnarray}
\Phi_P(k)&=&\frac{i}{3}\int\,\frac{d^4q}{(2\pi)^4}\,G_{\mu\nu}(q-k)\,
\frac{1}{2}{\rm tr}\Biggl[\gamma_5\,\Gamma^{\nu}(k,q)\,S_d(q)\,\gamma_5
\Phi_P(q)\,S_u(q)\,\gamma_{\mu}\label{bschir1}\\
&&+\gamma_5\,\gamma_{\nu}\,S_d(q)\,\gamma_5\Phi_P(q)\,S_u(q)\,
\Gamma^{\mu}(q,k)\Biggr].\nonumber
\end{eqnarray}
 Commuting $\gamma_5$ through propagators and noting 
that in the chiral limit,
\begin{eqnarray}
S_d(-q)S_u(q)&=&-\frac{1}{A^2(q)q^2-B^2(q)},\nonumber
\end{eqnarray}
Eq.~\ref{bschir1} can be rewritten as
\begin{eqnarray}
\Phi_P(k)&=&\frac{i}{3}\int\,\frac{d^4q}{(2\pi)^4}\,G^{\mu\nu}(q-k)\,
\frac{\Phi_P(q)}{A^2(q)q^2-B^2(q)}\frac{{\rm tr}\Bigl[(\Gamma_{\mu}(q,k)
-\gamma_5
\Gamma_{\mu}(k,q)\gamma_5)\gamma_{\nu}\Bigr]}{2}
\label{phifunct}
\end{eqnarray}
To ensure that $B(k)$ and $\Phi_P(k)$ satisfy the same equation in the chiral 
limit, therefore producing a massless pion, the following condition should be 
satisfied 
\begin{eqnarray}
{\rm tr}\Bigl[\Gamma_{\mu}(q,k)\,(1+\frac{A(k)}{B(k)}\ \! / \! \! \! k)\,
\gamma_{\nu}\Bigr]&=&\frac{{\rm tr}\Bigl[(\Gamma_{\mu}(q,k)+
\Gamma_{\mu}(k,q))\gamma_{\nu}]}{2}\label{chirconstr}
\end{eqnarray}
Clearly, the commonly used bare quark gluon vertex, $\Gamma_{\mu}=
\gamma_{\mu}$ satisfies this condition. Furthermore, the condition 
Eq.~\ref{chirconstr} is 
also satisfied by the following part of the Ball-Chiu vertex 
\begin{eqnarray}
\Gamma^{\mu}(p',p)&=&\frac{A(p')+A(p)}{2}\gamma^{\mu}+\frac{(p'+p)^{\mu}}
{p'^2-p^2}\frac{A(p')-A(p)}{2}\,(\ \! / \! \! \! p'+\ \!
 / \! \! \! p).\label{gamchir}
\end{eqnarray}
The $B(p')-B(p)$ part of the Ball-Chiu vertex would have 
contributed to the lefthandside of Eq.~\ref{chirconstr} while it does not 
contribute to the righthandside. The contribution of this term is small 
compared to the sum of other two terms. Therefore, it is a good approximation 
to drop this term. The vertex Eq.~\ref{gamchir} enables us to preserve the 
chiral limit while including the dominant part of the quark-gluon vertex 
dressing as suggested by the Ward-Takahashi identity. This approximate 
form(Eq.~\ref{gamchir}) of the Ball-Chiu vertex is used for the 
{\em quark-gluon coupling} in this work. 

Since the functions $B(k)$ and 
$\Phi_P(k)$ satisfy the same equation in the chiral limit, they are equal upto
a proportionality constant. With the help of the BS normalization 
condition~\ref{bsnorm} it is found~\cite{JOHNSON,DELSCAD} that
\begin{equation}
\Phi_P(k)=\frac{1}{f_{\pi}}B(k),
\end{equation}
where $f_{\pi}$ is the pion decay constant.

In order to complete the description of the model, one needs to choose a model
for the gluon propagator $G^{\mu\nu}(q)$. The choice of $G^{\mu\nu}(q)$ has 
been discussed in various papers\cite{JAINMUN2}. Usually, the 
ultraviolet$(q\rightarrow\infty)$ or asymptotic behavior of $G^{\mu\nu}(q)$ is
borrowed from QCD calculations, while its infrared$(q\rightarrow 0)$ or 
confining behavior is given by a sharply falling function such as $1/q^4$ or
$\delta^4(q)$, to incorporate confinement. The problem with $1/q^4$ behavior 
is that it is not an integrable singularity and one needs to introduce an 
infrared cutoff. While the $\delta^4(q)$ form does not have this 
problem, it is perhaps too simple a form to represent the physics in the 
infrared region; this issue requires further study. {\bf In this paper, we 
show that an infrared finite propagator can not be ruled out on the basis of 
quark Dyson-Schwinger and Bethe-Salpeter equations}. The model used here is 
\begin{equation}
G^{\mu\nu}(q)=(g^{\mu\nu}-\frac{q^{\mu}q^{\nu}}{q^2})\,[\,G_{{\rm IR}}+
\,G_{{\rm UV}}],
\end{equation}
where the infrared($q\rightarrow 0$) behavior, $G_{{\rm IR}},$ is modeled by a 
finite term
\begin{equation}
G_{{\rm IR}}(q)=G\,e^{-q^2/\sigma^2},
\label{infrared}
\end{equation}
while the asymptotic($q\rightarrow \infty$) form, $G_{{\rm UV}}$, is taken 
from 
perturbative QCD calculations
\begin{equation}
G_{{\rm UV}}(q)=\,2\pi^2\frac{d}{(q^2+\sigma^2)\,\ln(\tau+q^2/
\Lambda_{QCD}^2)} \ ,  \label{ultraviolet}
\end{equation}
with $d=12/(33-2\,N_f)=\frac{4}{9}$, where $N_f=3$ is the number of flavors. 
The QCD scale parameter $\Lambda_{QCD}$, determined by fitting high energy 
experiments(Particle Data Group, 1990), is chosen to be 225 MeV. The constant 
$\tau$ ensures the positivity of the asymptotic piece as $q\rightarrow 0$, and
results are not very sensitive to this parameter. The $\tau$ is chosen to be 
$\tau=3$. The $\sigma^2$ in the denominator of the UV piece is introduced to 
ensure that the infrared piece is the dominant contribution at low energies. 
Similar forms for the dressed gluon propagator have been used in the 
literature\cite{JAINMUN1,ROBERTS5,TANDY1,TANDY2} with considerable success in 
preliminary applications of Dyson-Schwinger Equations to hadronic physics. 
Here, we develop this approach using a dressed quark-gluon vertex while 
maintaining the chiral limit, {\bf and show that quarks can be confined with 
an infrared finite interaction}. Aside from current quark masses, $m_{u,d}
\approx 6\pm 2$ 
MeV and $m_s\approx 150 \pm 50$ MeV~\cite{GASSERLEUT}, which are also the 
input parameters of QCD, there are only {\bf two} unconstrained 
parameters($G,\sigma$) to vary to predict the data. Parameters 
$G=1.97 10^{-4}{\rm MeV}^{-2}$ and $\sigma=750$ MeV are chosen to give the 
optimum overall fit. We choose the current quark mass of the strange quark to 
fine tune the kaon mass. The current quark masses used in the calculation are:
$m_{u,d}=3$ MeV, and $m_s=60$ MeV. The ratio $m_s/m_u$ is well within  
acceptable limits~\cite{GASSERLEUT}. In the next section, we discuss the 
solution of the quark Dyson-Schwinger equation.

\section{QUARK PROPAGATORS AND CONFINEMENT}   
\label{section3}             
The numerical solution of the DS equation Eq.~\ref{sd1} is performed through 
iteration to find the quark propagator functions $A(p^2)$ and $B(p^2)$ in 
Euclidean metric. The 
details of the numerical methods are explained in the Appendix.
\begin{figure}[thb]
\begin{center}
\mbox{
   \epsfxsize=3.4in
   \epsfysize=3.0in 
\epsffile{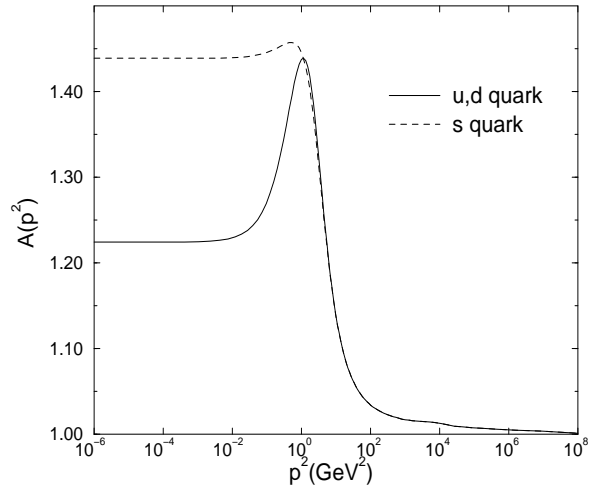}
}
\end{center}
\caption{The quark propagator function $A(p)$ is shown for up/down, and 
strange quarks. Dressing of the quark-gluon vertex causes the peak observed in
 $A(p)$.}
\label{auas.fig}
\end{figure}
\begin{figure}[thb]
\begin{center}
\mbox{
   \epsfxsize=3.4in
   \epsfysize=3.0in 
\epsffile{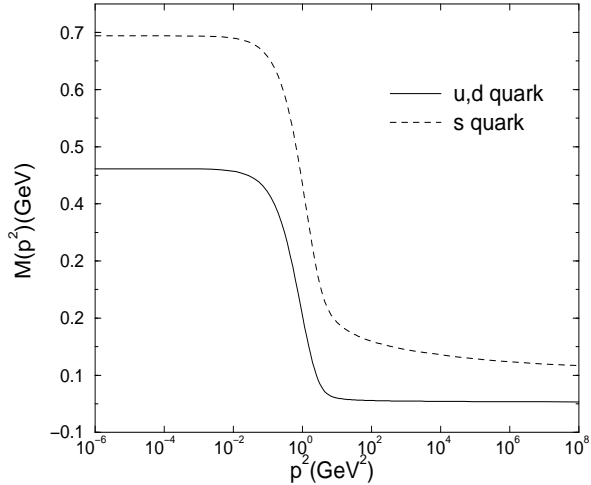}
}
\end{center}
\caption{Quark mass functions $M(p)$ are shown for up/down, and strange 
quarks. At the origin($p^2=0$), quark masses are closer to constituent quark 
mass values used in nonrelativistic quark models. Asymptotically, quark mass 
values approach current quark mass values.}
\label{mums.fig}
\end{figure}
Solutions for $A(p^2)$ and $M(p^2)\equiv B(p^2)/A(p^2)$ in the spacelike 
region are shown in Figures \ref{auas.fig} and \ref{mums.fig}.
According to the results in Figures \ref{auas.fig} and \ref{mums.fig}, for low
momenta ($p<1$ GeV) quark masses are close to those of the constituent quarks; 
whereas, as momentum increases, quarks start behaving as current 
quarks($A_f(q^2) \rightarrow 1,B_f(q^2) \rightarrow m_0$). In the infrared 
region, the quark masses approach the constituent mass values. The behavior of
 function $A_f(q^2)$ is different than that of $M_f(p^2)$; $A_f(q^2)$ reaches 
 its maximum value at intermediate energies, where the scale is given by 
$\sigma$. This is a different behavior than results presented in 
Ref.~\cite{JAINMUN2}, for the bare quark-gluon coupling case. In their 
results, both $M(p^2)$ and $A_f(q^2)$ are monotonically decreasing functions. 
In our case, the dressing of the quark-gluon vertex gives rise to the 
nonmonotonic behavior we found for $A_f(q^2)$. As one increases the coupling 
strength G, monotonic behavior of $A$ is restored. 
\subsection{Test of Confinement}
\begin{figure}
\begin{center}
\mbox{
   \epsfxsize=3.4in
   \epsfysize=3.0in 
\epsffile{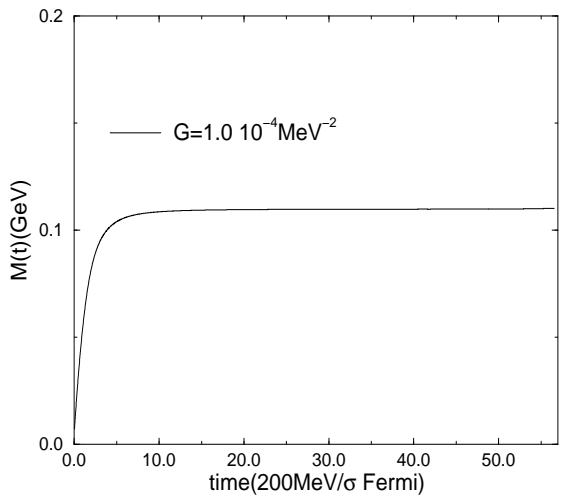}
}
\end{center}
\caption{Mass pole as a function of time is shown. Asymptotically, $M(t)$ 
approaches a constant, which indicates a real mass pole(unconfined  quark).}
\label{moft1.plt}
\end{figure}

\begin{figure}
\begin{center}
\mbox{
   \epsfxsize=3.4in
   \epsfysize=3.0in 
\epsffile{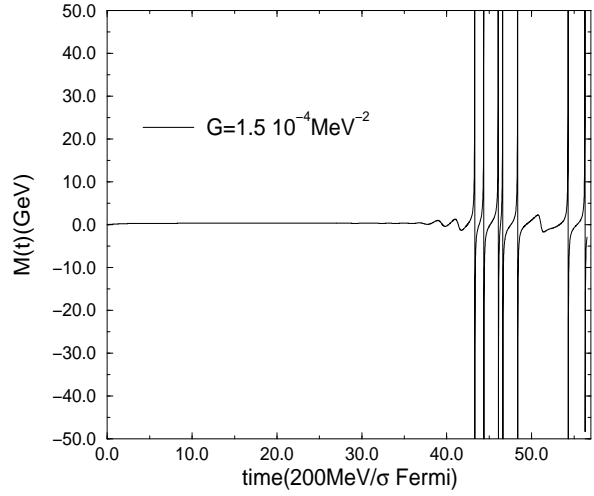}
}
\end{center}
\caption{Mass pole as a function of time is shown. Oscillations indicate that 
quark mass pole is complex.}
\label{moft1.5.plt}
\end{figure}

\begin{figure}
\begin{center}
\mbox{
   \epsfxsize=3.4in
   \epsfysize=3.0in 
\epsffile{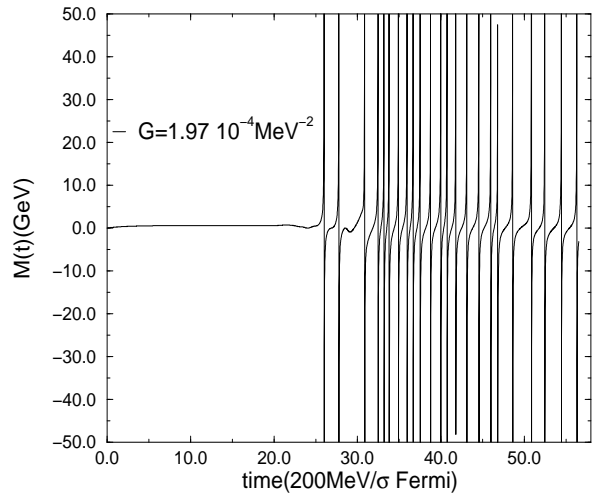}
}
\end{center}
\caption{As the strength of the coupling, $G$, is increased $M(t)\propto {\rm 
tan}(t)$ behavior clearly sets in.}
\label{moft1.97.plt}
\end{figure}

Confinement is the property that only color singlet hadrons are observed in 
nature. In QCD, confinement is obtained dynamically due to the nonabelian, 
hence self-interacting, nature of gluons.  A natural result of 
confinement is that no free quark state should be observed(a free quark state 
has a net color.) In QFT, an {\em n-body bound state} is defined by the pole 
of the n-body propagator. The familiar 2-body(Bethe-Salpeter, Gross) and 
3-body(Faddeev) equations are obtained, based on this definition, by looking 
for the poles in the two and three body propagators. Similarly, it is natural 
to expect that a 1-body(or {\em free}) state should be identified
by the pole of the one body propagator. Therefore, if quarks are confined, 
quark propagators should not have poles in the timelike$(p^2>0)$ 
region.\footnote{There is an alternative to this approach, which is developed 
within the context of the Gross equation in Ref.~\cite{GROSS}. In that 
approach, quarks are allowed 
to be on shell as long as they are in the vicinity of off-shell quarks, and 
confinement is realized through a relativistic generalization of the linear 
potential.},  
because such a pole permits an asymptotically(in space) free quark wave to 
exist. The absence of poles in the quark propagators is, however, not 
conclusive evidence for confinement. In order to be able to claim that a 
theory is confining, one has to also show that a diquark, or any other color 
nonsinglet stable bound state does not exist. Here, we restrict our 
discussion of confinement to quarks only. Therefore from here on, 
``confinement'' refers to ``lack of free quarks'' rather than 
the more general definition of ``lack of colored states.'' In order to test 
whether a quark propagator,  given in Euclidean metric, leads 
to confinement, one needs a procedure to determine the presence of any 
poles in the timelike region.\footnote{An alternative realization of 
confinement can be obtained by simply defining the quark mass function such 
that quark can never be on shell.\cite{KREWALD} This approach amounts to 
having quark mass function $M(p^2)$ as input and the effective gluon 
propagator $G^{\mu\nu}(q)$ as unknown in the quark Dyson-Schwinger equation. 
In this approach, a unique determination of the gluon propagator is not 
possible. Therefore, one is forced to make a separable interaction 
approximation.}  For simple cases such as a free fermion, the Euclidean 
expression for the propagator can be readily used to see if there are any 
poles when the propagator is continued to the Minkowski metric, i.e.
\begin{equation}
\frac{1}{p_E^2+m^2}\ \rightarrow\ \frac{1}{p^2-m^2},
\end{equation}
where there is a pole at $p^2=m^2$. For the dressed quark 
propagator, the confinement test question is: does
\begin{equation}
\frac{B(p_E^2)}{A^2(p_E^2)p_E^2+B^2(p_E^2) }  
\end{equation}
have a pole? For this test, the procedure used in Ref.~\cite{HRW} is adopted 
to determine whether a quark propagator given in Euclidean metric has poles, 
when continued to Minkowski metric. The starting point is the 
definition of a generalized quark propagator 
\begin{equation}
S(p)\equiv \int d\mu\,\rho(\mu)\,\frac{i}{\ \! / \! \! \! p-\mu}\ ,
\end{equation}
where $\rho(\mu)$ is a spectral density function. The Euclidean metric 
expression for ${\rm tr}[S_E(p)]$ is
\begin{equation}
{\rm tr}[S_E(p)]=4\,\int d\mu\,\rho(\mu)\,\frac{\mu}{p^2+\mu^2}\  .
\end{equation}
Using contour integration methods, the Fourier transform of this trace is 
found to be
\begin{eqnarray}
\Delta(t)&=&\int \frac{dp}{2\pi}\,e^{ipt}\,{\rm tr}[S_E(p)]\\
&=&2\int d\mu\, \rho(\mu)\, e^{-\mu t}\ . \label{Delta}
\end{eqnarray}
We define $\bar{\mu}(t)$, the average effective mass(pole location) of the 
propagator $S(p)$ as a function of time by  
\begin{eqnarray}
\bar{\mu}(t)&\equiv&-\frac{\partial}{\partial t} \ln(\Delta(t))\\
&=&\frac{\int d\mu\, \rho(\mu)\,\mu\, e^{-\mu t}}{\int d\mu\, 
\rho(\mu)\, e^{-\mu t}}\ .
\end{eqnarray}
Let us assume that $S(p)$ has at least one pole for a finite $p^2>0$. Let 
$\rho(\mu)=\sum_{i=0}^{n-1}c_i\,\delta(\mu-m_i)$, where $n\geq 1$ and 
$m_i<m_{i+1}$. This assumption leads to a discrete 
average
\begin{equation}
\bar{\mu}(t)=\frac{\sum_{i=0}^{n-1}c_i\,m_i\,e^{-m_it}}{
\sum_{i=0}^{n-1}c_i\,e^{-m_it}}.
\end{equation}
Taking the limit of this expression at infinite time $t\rightarrow \infty$, 
one
has
\begin{equation}
\lim_{t\rightarrow \infty} \bar{\mu}(t)\cong m_0.
\end{equation}
Therefore, this averaging procedure gives the smallest pole of the propagator 
$S(p)$.
If this limit exists and it is real, then there is at least one finite pole 
for timelike momenta in the Minkowski metric, and therefore the propagator 
does not represent a spatially confined particle. On the other hand, if there 
is no finite limit or the limit is complex, then the propagator represents a 
confined particle, since the propagator does not have a real finite pole. This 
test for confinement has been applied to the quark propagator obtained from 
the numerical solution of the Dyson-Schwinger equation. There are three 
possible pole structures  for the quark propagator. These possibilities are
exemplified by the following:
\begin{itemize}
\item Real pole:
\end{itemize}
\begin{eqnarray}
S_E(p_E)&=&\frac{1}{p_E^2+m^2},\nonumber\\
\Delta(t)&\propto& e^{-mt},\nonumber\\
\bar{\mu}(t)&=&m,\nonumber
\end{eqnarray}
where the test produces the pole location as expected.
\begin{itemize}
\item Absence of poles:
\end{itemize}
\begin{eqnarray}
S_E(p_E)&=&e^{-p_E^2/(2\sigma^2)},\nonumber\\
\Delta(t)&\propto& e^{-\sigma^2t^2/2},\nonumber\\
\bar{\mu}(t)&=&\sigma t\rightarrow \infty,\nonumber
\end{eqnarray}
In this example, there is no finite pole and the propagator represents a 
confined 
particle. This analytic form is the same as the infrared piece of the 
effective gluon propagator~\ref{infrared}. 
\begin{itemize}
\item Complex poles:
\end{itemize}
\begin{eqnarray}
S_E(p_E)&=&\frac{1}{(p_E-ia)^2-m^2}\ , \nonumber
\end{eqnarray}
Here the pole is complex in general and purely imaginary when $a=0$. For this 
case, the test results gives analytically 
\begin{eqnarray}
\Delta(t)&\propto& e^{-at}{\rm sin}(mt),\nonumber\\
\bar{\mu}(t)&=&a+m\,{\rm tan}(mt-\frac{\pi}{2}).\nonumber
\end{eqnarray}
Therefore, the signature of complex poles appears as $M(t)\equiv\bar{\mu}(t)
\propto{\rm tan}(mt)$ behavior as $t\rightarrow \infty$, where the frequency 
of oscillations is proportional to the imaginary part of the quark mass pole. 
This is exactly the type of behavior found by applying the 
confinement test to the quark propagator obtained by numerically solving the 
DS equation~\ref{sd1}. Since the quark propagator is known only numerically, 
application of the test is numerical and details of the numerical methods are 
explained in the Appendix. Here we present the test results for three 
different coupling strengths, namely $G=1\times 10^{-4},1.5\times 10^{-4},$ 
and $1.97\times 10^{-4}{\rm MeV}^{-2}$. The
first case, $G=1\times 10^{-4}{\rm MeV}^{-2}$ is shown in 
Figure~\ref{moft1.plt}. 
According to this result, the pole location is finite($\approx$ 110MeV) and 
real. Therefore, this quark propagator does not represent a confined particle.
On the other hand, this is a case where the coupling constant is very small. 
As one increases the strength of the coupling to 
$G=1.5\,10^{-4}{\rm MeV}^{-2}$ an irregular oscillatory behavior 
sets in(Fig.~\ref{moft1.5.plt}). If the coupling strength is further increased
to $G=1.97\,10^{-4}{\rm MeV}^{-2}$, which is the parameter that is used to fit
all observables in this paper, {\bf the oscillations clearly displays the 
${\rm tan}(mt)$ behavior(Fig.~\ref{moft1.97.plt}) which indicates that quarks 
have complex mass poles}. According to this result(Fig.~\ref{moft1.97.plt}), 
{\bf the average distance a quark can travel before it hadronizes}, which is 
given by the average distance between the peaks(singularities) of the  $M(t)$ 
function, is approximately $D\times 200{\rm MeV}/\sigma\,{\rm Fermi}=1.45
\times 200/750={\bf 0.39\,{\bf Fermi}}$, where $D\approx 1.45$ is the average 
spacing between the peaks. This result is in very good agreement when compared
with the sizes of various hadronic bound states such as the pion($r_{\pi}=.66$ 
Fermi) and the kaon($r_K=.53$ Fermi).        
A recent study~\cite{CONRAD} of the DS equation within the context of QED in 
three dimensions similarly finds complex conjugate mass poles in the fermion 
propagator.   
It is important to emphasize that the analytic form of the gluon propagator
used in this calculation is not the only possible choice to produce confined
quarks. In fact, we have obtained similar results with other infrared singular 
gluon propagators. Therefore, it is not possible to single out a specific 
analytic form for the gluon propagator solely on the basis of the confinement 
test. Just as there are infinitely many confining potentials in 
nonrelativistic quantum mechanics, there are infinitely many confining 
effective gluon propagators in field theory. Therefore, additional tests such 
as the prediction of meson observables are needed to determine if the gluon 
propagator ansatz makes physical sense.

Having shown that it is possible to obtain confined quarks using a gluon 
propagator with a gaussian type of infrared behavior, we now turn to the 
quark-antiquark bound state problem.
  
\section{QUARK-ANTIQUARK BOUND STATES}
\label{section4}
\begin{figure}
\begin{center}
\mbox{
   \epsfxsize=3.4in
   \epsfysize=3.0in 
\epsffile{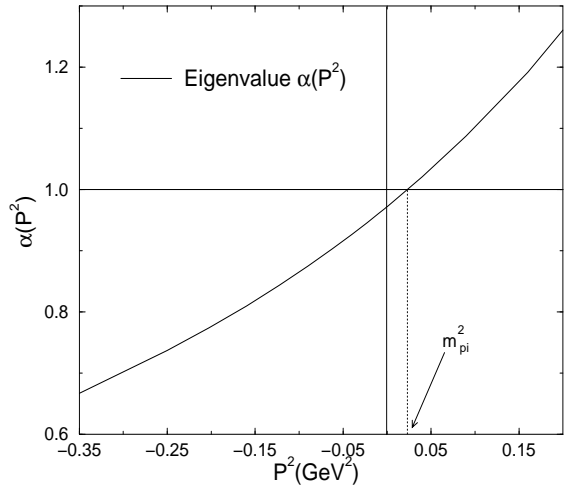}
}
\end{center}
\caption{ Dependence of eigenvalue $\alpha$ on $P^2=m_{\pi}^2$. Location of 
the bound state momentum is given by the intersection defined by 
$\alpha=1=f(P^2)$.}
\label{pion.alpha.plt}
\end{figure}

\begin{figure}
\begin{center}
\mbox{
   \epsfxsize=3.4in
   \epsfysize=3.0in 
\epsffile{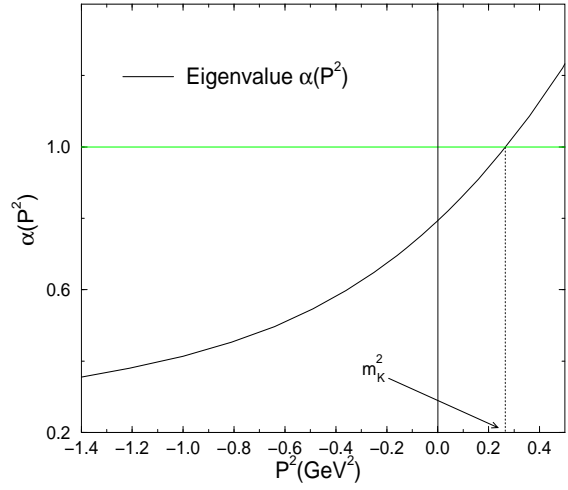}
}
\end{center}
\caption{ Determination of kaon mass. }
\label{kaon.alpha.plt}
\end{figure}

Before we embark on solving the BS equation, it is necessary to clarify 
a technical problem. In the previous section, dressed quark propagators were 
calculated in the Euclidean metric(or spacelike momentum region). For the 
Bethe-Salpeter equation, usage of the Euclidean metric is more
problematic.\footnote{Recent efforts to solve the BS 
equation directly in Minkowski metric are given in 
Refs.~\cite{WILLIAMS1,WILLIAMS2}}  Unlike the quark Dyson-Schwinger
equation, the 
Bethe-Salpeter equation involves the external total bound state momentum $P$, 
which has to eventually represent a physical particle(bound state) with a real
positive mass. Therefore, the four momentum of the particle should be 
$P=(m,\vec{0})$, for which $P^2=m^2>0$ represents a timelike particle. It 
follows that, in order to be able to perform the integrations in 
Eq.~\ref{bseq1} in Euclidean metric, one needs to know the functions 
$A(q_+^2),B(q_+^2)$ for $q_+^2=m^2\eta_1\eta_2-q^2+i\,mq_0$, where 
$\eta_{1,2}$ is chosen to be
\begin{equation}
\eta_{1,2}=\frac{m_{1,2}}{m_1+m_2},
\end{equation}
and $m_{1,2}\equiv M_{u,d}(0)$. On the other hand, functions 
$A(q_+^2),B(q_+^2)$ are known only for real and spacelike $q_+^2$. 
\begin{figure}
\begin{center}
\mbox{
   \epsfxsize=2.9in
   \epsfysize=1.8in 
\epsffile{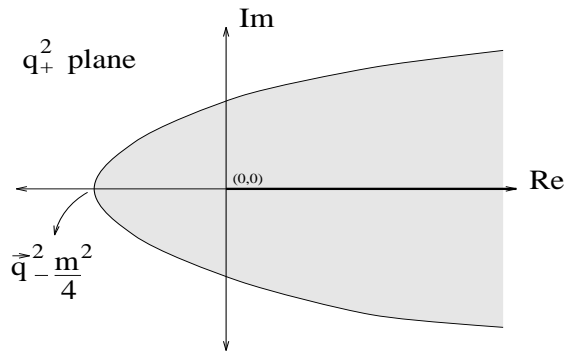}
}
\end{center}
\caption{Argument domain of functions $A(q_+^2)$ and $B(q_+^2)$ for physical
bound states($m^2>0$). The positive axis shows the spacelike $q_+^2$ region. }
\label{cmplx.fig}
\end{figure}
The thick line in Fig.~\ref{cmplx.fig}(the positive $q_+^2$ axis) is 
where functions $A(q^2)$ and $B(q^2)$ has been calculated by solving the 
Dyson-Schwinger equation. The domain where these functions are needed is shown
by the shaded region in Fig.~\ref{cmplx.fig}. At this point, there are three 
options. The first one, used in Ref.~\cite{ROBERTS7} among other works, is 
to assume an analytic functional form that fits the numerical functions 
$A(q^2),B(q^2)$ on the positive real axis. Once this assumption is 
made one can use these analytic functions over the entire complex plane. The 
second approach, which has been used in Ref.~\cite{JAINMUN1}, is to Taylor 
expand the functions $A(q^2),B(q^2)$ around their real values to extend the 
results to the complex plane. Both of these methods have drawbacks. It has 
been shown in previous section that the quark propagator has complex conjugate
mass poles in the spacelike region.   It has also 
been shown\cite{STAINSBYCAHILL} that functions $A(q^2),B(q^2)$ possibly
have poles and branch cuts in the complex plane, which complicates the above 
methods. Therefore sampling the momentum values away from the spacelike 
momentum axis is problematic. We follow a third approach, which is also used 
in  lattice gauge theory calculations. In this approach, the bound state 
problem is solved for a set of spacelike bound state momenta, $P^2=-m^2<0$, 
which transforms the problem to Euclidean metric, thereby avoiding the complex 
argument problem, and the final results are then extrapolated back to 
the physical region, $P^2>0$. Since it is the final results such as form 
factors and decay constants that are extrapolated ( rather than functions 
$A(q^2), B(q^2),$ which are integrated out in calculations of observables),  
this method has the benefit of explicitly displaying the reliability of the 
extrapolation. The only assumption is the analyticity of the {\em observables}
as a function of the bound state mass. {\bf In this method, there is no need 
to assume that functions $A(q^2),B(q^2)$ are analytic}. The procedure for the 
solution is as follows: First the 
Bethe-Salpeter equation~\ref{bseq1} is discretized\footnote{see Appendix for
details of our numerical techniques.} and transformed into a matrix 
Equation
\begin{equation}
H_{P^2}\,\Phi=\Phi,
\label{eigenvalue1}
\end{equation}
where $P^2=m^2>0$ is the bound state mass and the implicit eigenvalue of this 
matrix equation. Since Eq.~\ref{eigenvalue1} will be solved for spacelike 
bound states $P^2=m_i^2<0$, one will not be able to find any solutions unless 
an artificial eigenvalue $\alpha_i$ is introduced to Eq.~\ref{eigenvalue1}
\begin{equation}
H_{m_i^2}\,\Phi=\alpha_i\Phi.
\label{eigenvalue}
\end{equation}
One proceeds by finding a set of solutions $\{m_i^2<0, \alpha_i\}$ to the 
above equation. This is done by using an inverse iteration technique as 
explained in the Appendix. Using the set of solutions $\{m_i^2<0, \alpha_i\}$,
 a 
functional relationship between $\alpha_i$ and $m_i^2$ can be established 
\begin{equation}
\alpha_i=f(m_i^2). 
\label{alphaf}
\end{equation}
It is only when $m_i^2=m^2>0$ and $\alpha_i=1$ that one recovers the original 
BS equation(\ref{eigenvalue1}). Therefore, location of the $m^2$ that gives 
$\alpha=1$ is the eigenvalue and mass of the physical bound state in question.
The most general form for the spin-space part of the BS vertex function for 
pseudoscalar mesons is given by
\begin{equation}
\chi_P(k)=i\gamma_5[\Phi_0+\ \! / \! \! \! \! P\,\Phi_1+\ \! / \! \! \! k\,
\Phi_2+[\ \! / \! \! \! k,\ \! / \! \! \! \! P]\,
\Phi_3].\label{pseudoscalarmode}
\end{equation}
In the pseudoscalar meson channel, the dominant contribution to the BS 
vertex function comes from the first term~\cite{JAINMUN1,ROBERTS5},
\begin{equation}
\chi_P(k)\approx i\gamma_5\Phi_P(k)
\label{pseudoscalarmode2}
\end{equation}
This dominance is not surprising
since the Dirac structure of the leading term in~\ref{pseudoscalarmode2}
is the same as that of pointlike pion-quark coupling. Therefore, we only 
consider the leading term in our analysis.\footnote{For vector mesons, the 
leading term might not be the only important one.}
The angular dependence of the BS vertex function is made explicit by 
expanding it in terms of Tchebyshev polynomials
\begin{equation}
\Phi_P(k)=\sum_{n=0}^{\infty}\Phi^n_P(|k|)T^n({\rm cos}\gamma).
\label{tchebexp}
\end{equation}
For the bound states considered in this work($\pi,\pi^*,{\rm K}^{\pm}$), the 
dominant contribution to $\Phi_P(k)$ comes from the $T_0$ polynomial. Test 
runs for cases where higher($n>0$) Tchebyshev polynomials 
are included showed that the contribution of the higher Tchebyshev 
polynomials are negligible, which is in agreement with conclusions in 
Ref.~\cite{JAINMUN1}.
After solving the BS equation numerically, a relationship~\ref{alphaf} between 
the largest eigenvalue\footnote{The second largest $\alpha$ leads to the first
 excited state.}  $\alpha$ and $m^2$ is constructed by using a polynomial fit
\begin{equation}
\alpha=\sum_{i=0}^{n} c_i(m^2)^i.
\end{equation}
It has been determined that for all of the bound states under consideration 
$n=3$ gave a satisfactory fit. 
The extrapolations done to find the pion and kaon masses are shown 
respectively in Fig.~\ref{pion.alpha.plt} and Fig.~\ref{kaon.alpha.plt}. 
\begin{figure}
\begin{center}
\mbox{
   \epsfxsize=3.4in
   \epsfysize=3.0in 
\epsffile{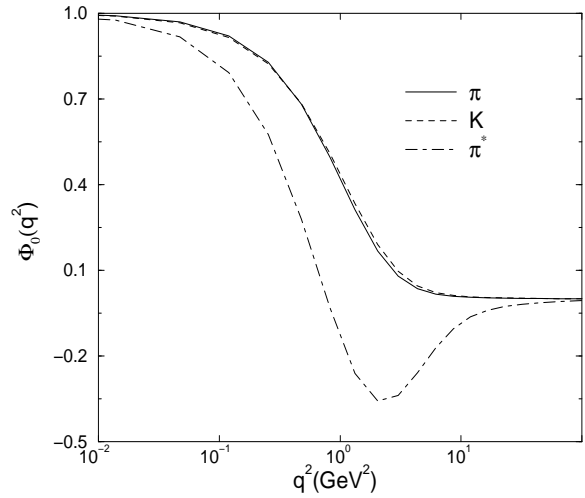}
}
\end{center}
\caption{The pion and kaon ground and pion first excited state BS 
wavefunctions $\Phi_{P}^0(|k|)$ are shown all for $P^2=0$.}
\label{pipistark.wfn.plt}
\end{figure}
The pion ground and first excited state and kaon groundstate wavefunctions
are shown in Fig.~\ref{pipistark.wfn.plt}. 
The first excited state of the pion has a node, while the ground state 
wavefunctions are positive definite. 
The kaon BS wavefunction is more spread out in momentum space than that of 
pion.

\subsection{The Electromagnetic form factor using Euclidean metric}
\begin{figure}
\begin{center}
\mbox{
   \epsfxsize=3.4in
   \epsfysize=3.0in 
\epsffile{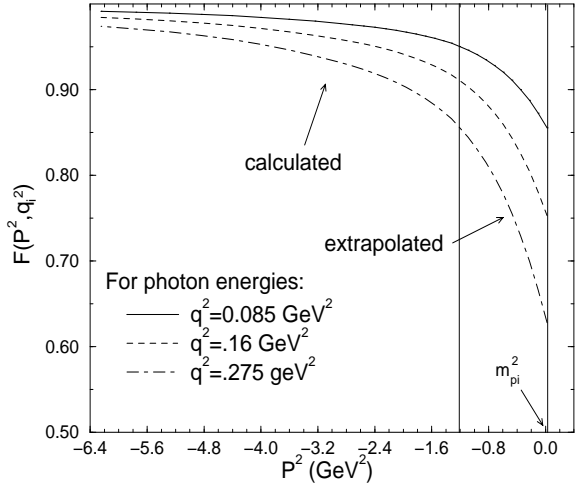}
}
\end{center}
\caption{These plots show how the form factor extrapolation is done for 
various photon energies. The extrapolated results are on the right hand
side(timelike, $P^2=m_{\pi}^2>0$) of bound state momentum axis. }
\label{extrap}
\end{figure}

\begin{figure}
\begin{center}
\mbox{
   \epsfxsize=3.4in
   \epsfysize=3.0in 
\epsffile{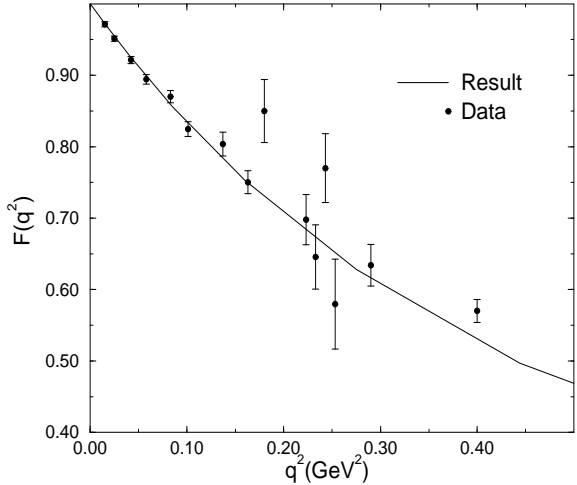}
}
\end{center}
\caption{Extrapolated result for electromagnetic pion form factor.}
\label{pion.formf.plt}
\end{figure}

\begin{figure}
\begin{center}
\mbox{
   \epsfxsize=3.4in
   \epsfysize=3.0in 
\epsffile{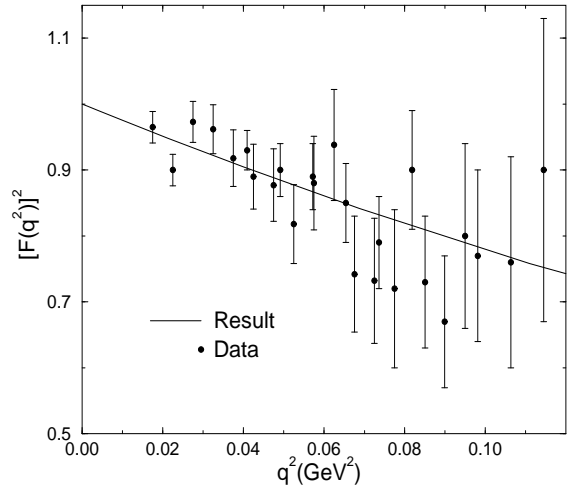}
}
\end{center}
\caption{ Electromagnetic form factor of Kaon.}
\label{kaon.formf.plt}
\end{figure}

Electromagnetic current conservation  
\begin{equation}
\partial_{\mu}J^{\mu}_{em}=0,
\end{equation}
implies that in momentum space one must have $q_{\mu}J^{\mu}_{em}=0$. Since 
both initial and final pion states are on their mass shells, we have 
$p^2=m^2=(p+q)^2$, or $2p\cdot q+q^2=0$. Therefore, the definition of the form
factor takes the following form
\begin{equation}
<\pi^+(p+q)|J^{\mu}_{em}(0)|\ \pi^+(p)>=F_{\pi}(q^2)(2p^{\mu}+q^{\mu}).
\end{equation}
\begin{figure}
\begin{center}
\mbox{
   \epsfxsize=3.0in
   \epsfysize=1.0in 
\epsffile{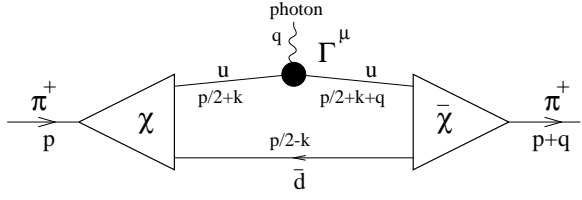}
}
\end{center}
\caption{Electromagnetic Form Factor of Pion in the impulse approximation(i.e.
interaction of the initial and final pion with each other is neglected.)}
\label{formf.fig}
\end{figure}
This matrix element is represented  by the Feynman diagram shown in 
Figure~\ref{formf.fig}.\footnote{Only the interaction of the photon with the u 
quark is shown. There is also 
a second diagram where the photon interacts with the $\bar{d}$ quark.}
As the photon momentum $q$ vanishes, the pion is perceived as a point charge. 
Therefore, it is expected that $F(0)=e$, and the charged pion has an 
electromagnetic charge of $e$. To arrive at this result, two conditions 
must be satisfied; namely, the BS wavefunction should be properly 
normalized, and the conservation of the electromagnetic current at the 
photon-quark interaction vertex should be taken into account. The dressed 
electromagnetic vertex function $\Gamma^{\mu}(p',p)$ 
for the photon quark coupling satisfies the Ward-Takahashi identity 
Eq.~\ref{wtid}, which is an indirect statement of the conservation of the 
electromagnetic current. For the {\em photon-quark} vertex we use the full 
Ball-Chiu vertex~\ref{BallChiu}.
In the Breit frame, the initial and final pion momenta are 
\begin{equation}
P_i=(i\sqrt{m_i^2-\frac{q^2}{4}},0,-\frac{q}{2}),\ \ P_f=(i\sqrt{m_i^2-
\frac{q^2}{4}},0,\frac{q}{2}).
\end{equation}
Therefore, $(P_i)^2=(P_f)^2=-m_i^2<0$ are both 
spacelike. Calculation of the form factor is done for a set of spacelike bound
state masses $m_i^2,\,(i=1\cdots 21)$ and the resultant set of form factor 
functions are extrapolated at each photon momentum. For example, the pion form 
factor at photon momentum of $q_i^2$ is
\begin{equation}
F(m^2,q_i^2)=\sum_{j=0}^{n} C^j(q_i^2)\, m^{2j},
\end{equation}
where $C^j(q_i^2)$ are the coefficients of a fit to the numerical result for 
$F$ for photon momentum of $q_i^2$. 
Therefore, the physical result for the pion form factor at the photon momentum
$q_i^2$ is $F(M_{\pi}^2,q_i^2)$.
In Figure~\ref{extrap}, we show the extrapolation for three different 
photon momenta, $q^2=0.085$ GeV$^2$, $q^2=0.16$ GeV$^2$, and $q^2=0.275$ 
GeV$^2$. For each case, the result of extrapolation is given by the value of 
the form factor functions at $P^2=m_{\pi}^2$. 
In order to ensure the reliability of the fit, a large number(21) of 
form factor calculations {\em at each photon momentum} has been performed. The
order of the polynomial fit used is $n=7$. According to the extrapolation 
results(Fig.~\ref{extrap}) as the photon momentum increases, the reliability 
of the extrapolation decreases. This 
difficulty is common to all solution methods(within the covariant DS-BS 
formalism) which use Euclidean metric.\footnote{See 
Refs.~\cite{KISSLINGER} for a discussion of the transition to 
asymptotic(large $q^2$) behavior of the pion form factor within a light-cone 
Bethe-Salpeter formalism.} The reason for this sensitivity is the increase
in the curvature of the function $F(m^2,q_i^2)$ with increasing $q_i^2$. We 
present the form factor calculation for 
pion(Fig.~\ref{pion.formf.plt}) and kaon(Fig.~\ref{kaon.formf.plt}) cases in 
the region where the extrapolation is reliable.  
The data points for the pion form factor are taken from 
Refs.~\cite{PION1,PION2}. 
Data available from earlier experiments~\cite{KAON1,KAON2} for the kaon form 
factor are  very poor. Experiments at  TJNAF will hopefully provide better 
measurements for both pion and kaon form factors. 
The pion decay constant, $f_{\pi}$, is defined by the vacuum to one pion 
matrix element of the axial vector current:  
\begin{equation}
<0|J_{5\mu}^i(x)|\ \pi^j(p)>\equiv if_{\pi}\delta^{ij}p_{\mu}e^{-ip\cdot x}.
\end{equation}
\begin{figure}
\begin{center}
\mbox{
   \epsfxsize=2.5in
   \epsfysize=0.8in 
\epsffile{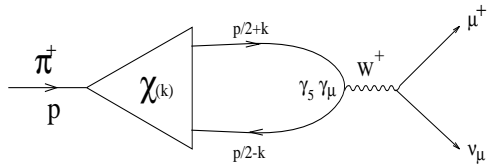}
}
\end{center}
\caption{Pion Decay}
\label{fpi.fig}
\end{figure}
For a $\pi^+$ meson at $x=0$, this definition becomes:
\begin{equation}
<0|\bar{\Psi}_d(0)\gamma_{\mu}\gamma_5\frac{\lambda^-}{2}\Psi_u(0)|\pi^+(p)>
\equiv 
if_{\pi}p_{\mu}.
\end{equation}
This matrix element corresponds to the Feynman diagram shown in 
Figure~\ref{fpi.fig}.
Calculation of decay constants are done using the same type of extrapolation 
technique used in the calculations of masses and form factors. Values found 
are $f_{K}=90.3(113)$ MeV, and $f_{\pi}=77.2(92.4)$ MeV, where numbers in 
parenthesis are the experimental measurements. The ratio of decay constants
$f_{K}/f_{\pi}=1.17(1.22)$ is in good agreement with the experimental 
value and comparable to those found in similar works. It is reported in 
Ref.~\cite{TANDY1} that next to leading order terms of the vertex 
function~\ref{pseudoscalarmode} contribute significantly to the decay 
constants. This is a possible explanation for the low values we find for the 
decay constants.  

\begin{figure}
\begin{center}
\mbox{
   \epsfxsize=2.4in
   \epsfysize=0.8in 
\epsffile{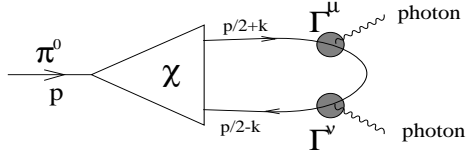}
}
\end{center}
\caption{ Neutral Pion Decay.}
\label{gpigg.fig}
\end{figure}
As a final application,  the neutral pion decay to two 
photons, $\pi\rightarrow \gamma + \gamma$ is considered. 
Neutral pion decay is of historical importance since it is 
associated with the {\em axial anomaly}.
The matrix element for $\pi^0\rightarrow\gamma\gamma$ 
decay(Fig.~\ref{gpigg.fig}) is
\begin{eqnarray}
{\cal T}(k_1,k_2)&=&-2i\frac{\alpha_{em}}{\pi f_{\pi}}\,\epsilon_{\mu\nu\rho
\sigma}\nonumber\\
&&\times\varepsilon^{\mu*}(k_1)\,\varepsilon^{\nu*}(k_2)\,k_{1\rho}\,k_{2
\sigma}\,M(k_1,k_2),
\end{eqnarray}
Since final photons are on-shell, $k_i^2=0$, and $P^2=(k_1+k_2)^2=2\,k_1\cdot 
k_2=m_{\pi}^2$. Therefore the scalar function  $M(k_1,k_2)$ is only a function
 of the pion mass. The decay rate 
\begin{equation}
\Gamma_{\pi^0\rightarrow\gamma\gamma}=\bigl(\frac{\alpha_{em}}{\pi f_{\pi}}
\bigr)^2\frac{m_{\pi}^2}{16\pi}\,M(m_{\pi}^2),
\end{equation}
is experimentally measured as $\Gamma_{\pi^0\rightarrow\gamma\gamma}=7.74\pm 
0.56$ eV, which means 
\begin{equation}
g_{\pi^0\gamma\gamma}\equiv M(m_{\pi}^2)=0.504\pm 0.019.
\end{equation}
It has been shown in Ref.~\cite{ROBERTS7} that, in chiral limit, irrespective 
of the details of quark propagators, as long as the photon-quark vertices are 
properly dressed to conserve the electromagnetic current and the BS vertex 
function is properly normalized, $g_{\pi^0\gamma\gamma}$ is analytically 
found to be 0.5. When the mass of the pion is taken into account, we find 
$g_{\pi^0\gamma\gamma}=.43$ which is close to the experimental value.  
\begin{table}
\centerline{
\begin{tabular}{|c|c|c|}
\hline
Observable & Calculated & Experimental \\ \hline\hline
$m_{\pi}$( MeV) & 148 & 139.6 \\ 
\hline
$f_{\pi}$( MeV) & 77.2  & 92.4 \\ 
\hline
$<r_{\pi}^2>^{1/2}$( Fermi) & .65 & .66\\  
\hline
$g_{\pi^0\gamma\gamma}$ & .43 & .504\\  
\hline
$m^*_{\pi}$( MeV) & 1245 & 1300$\pm$ 100\\  
\hline
$m_{K}$( MeV)& 515 & 495\\ 
\hline
$f_{K}$(MeV)& 90.3 & 113 \\ 
\hline
$<r_{K}^2>^{1/2}$( Fermi) & .54 &.53 \\ 
\hline
$f_{K}/f_{\pi}$ & 1.17 & 1.22 \\ 
\hline
$r_{K}/r_{\pi}$ & 0.83 & 0.8\\ 
\hline
\end{tabular}}
\caption{Summary of results}
\label{results}
\end{table}

A summary of the observables we calculated is given in Table~\ref{results}.
Error bars in experimental measurements are negligible unless indicated.

\section{CONCLUSION}
\label{section5}
In this work, we analyzed three aspects of the Dyson-Schwinger Bethe-Salpeter
equation approach. We have shown that it is possible to dress the quark-gluon 
vertex beyond the simple $\gamma^{\mu}$ form while maintaining the chiral 
limit condition. We have shown that an infrared finite gluon 
propagator can lead to confined quarks through generation of complex quark 
masses. It was found that, according to the model presented here, quarks can 
freely propagate only $\approx 0.4$ Fermi which is in very good agreement with
the hadronic bound state sizes. We have calculated all observables, including 
the pion form factor, using a Euclidean metric approach without relying on the 
analyticity properties of quark propagator functions $A(p^2)$ and $B(p^2)$. It
is found that the extrapolations associated with the usage of Euclidean metric
are reliable up to 1 ${\rm GeV}^2$ for calculation of masses, and up to around 
$.5 {\rm GeV}^2$ for the calculation of form factors. 
  
\section{ACKNOWLEDGMENTS}
We gratefully acknowledge very useful and encouraging discussions with Dr. 
Craig Roberts. Dr. Conrad Burden is gratefully acknowledged for his input. 
This research was supported, in part by the U.S. National 
Science Foundation international Grant INT9021617.  
One of us(\c{C}.~\c{S}.) has been supported by an
Andrew Mellon Predoctoral Fellowship at the University of Pittsburgh, and by 
the DOE through grant No. DE-FG05-88ER40435. We also acknowledge receipt of a 
grant of Pittsburgh Supercomputer Center computer time.
\appendix
\section{Numerical methods}
Solutions of integral equations are performed by first discretizing the 
integrals 
\begin{equation}
\int dq\, f(q) \longrightarrow \sum_{i=1}^n w_i\, f(q_i),
\end{equation}
where $w_i$ are integration weights for grid points $q_i$. In order to map 
the grid points and weights from interval $(-1,1)$ to $(0,\infty)$ we use the
arctangent mapping(Ref.~\cite{THK,TK})
\begin{equation}
y(x)=R_{min}+\frac{R_{d}{\rm tan}(\frac{\pi}{4}(1+x))}{1+\frac{R_{d}}{R_{max}
-R_{min}}{\rm tan}(\frac{\pi}{4}(1+x))},
\label{mapping}
\end{equation}
where
\begin{equation}
R_{d}=\frac{R_{med}-R_{min}}{R_{max}-R_{med}}(R_{max}-R_{min}).
\end{equation}
It follows that
\begin{equation}
y(-1)=R_{min},\,\,y(0)=R_{med},\,\,y(1)=R_{max}.
\end{equation}
Therefore, one can safely control the range$(R_{min},R_{max})$ and 
distribution$
(R_{med})$ of grid  points. With this discretization procedure, continuous 
integral equations are transformed into 
nonsingular matrix equations.
\subsection{Dyson-Schwinger Equation}
The DS equation involves two unknown functions $A$,$B$ which appear in two 
coupled equations(Eq.~\ref{sd1}). After discretizing the associated 
integrals, one has the following matrix equations
\begin{eqnarray}
B&=&\mu I+G_1\,F_1,\nonumber\\
A&=&I+G_2\,F_2.\label{coupled}
\end{eqnarray}
$A$,$B$,$F_1$, and $F_2$ are $n$ dimensional vectors where 
\begin{eqnarray}
F_1(i)&\equiv&B(i)/(B(i)^2 q_i^2+A(i)^2),\nonumber\\
F_2(i)&\equiv&A(i)/(B(i)^2 q_i^2+A(i)^2),\nonumber
\end{eqnarray}
and $G_1$ and $G_2$ are $n\times n$ matrices.
Coupled equations(Eq.~\ref{coupled}) are solved for $A,B$ by forward 
iteration. An arbitrary initial 
guess for functions(vectors) $A$ and $B$ is entered on the right hand side and 
the resulting vectors are iteratively used for the same process until a stable 
solution is achieved. Grid points in momentum space have been chosen for 
momenta between $R_{min}=0$ MeV and $R_{max}=10^5\sigma$ MeV where $\sigma$ is 
the relevant momentum scale of the problem. The median of the grid point 
distribution was $R_{med}=5\sigma$ MeV. This uneven distribution of grid 
points ensures that the concentration of grid points for lower 
momenta, that is where the integrand is maximum, is higher. Due to the smooth 
nature of $A$ and $B$ functions only 40 grid points sufficed to find stable
solutions. The number of iterations needed to find a stable result is around 
20.   
\subsection{Bethe-Salpeter Equation: Inverse Iteration Method}
Here we outline the inverse iteration method originally developed in 
Refs.~\cite{THK,TK}.

The BS equation can be brought into the following form
\begin{equation}
[H_{M^2}-\alpha]\Phi=0,
\label{matrixeq}
\end{equation}
One starts with an arbitrary vector $\chi^0$
\begin{equation}
\chi^0=\sum_{i=1}^N c_i\,\Phi_i
\end{equation}
where $\Phi_i$, $i=1..N$, satisfy
\begin{equation}
[H_{M^2}-\omega_i]\Phi_i=0,
\end{equation}
where $\omega_i,\,i=1..N$ are eigenvalues of the $H_{M^2}$ matrix. Next, an 
arbitrary first guess for $\alpha$ is chosen. It should 
be emphasized that eigenvalues which are not equal to $\alpha$ 
have no physical meaning, for they do not correspond to a solution of the BS 
equation(Eq.~\ref{matrixeq}).
In order to see if the initial guess $\alpha$ corresponds to one of the 
eigenvalues $\omega_i$, we construct the operator
\begin{equation}
K=\frac{1}{H_{M^2}-\alpha}.
\end{equation}
Operating $K$ on state $\chi^0$ n times produces
\begin{equation}
\chi^n=K^n\,\chi^0=\sum_{i=1}^N\,\frac{c_i}{(\omega_i-\alpha)^n}\,\Phi_i.
\end{equation}
When the number of iterations $n$ is sufficiently large(usually around ten), 
the dominant contribution to $\chi^n$ comes from the eigenvector $\Phi_j$
whose eigenvalue $\omega_j$ satisfies $|\omega_j-\alpha|<|\omega_i-\alpha|$ 
for all $i=1..j-1,j+1..N$. Therefore,
\begin{eqnarray}
\chi^{n}&\approx& \frac{c_j}{(\omega_j-\alpha)^n}\,\Phi_j,\nonumber\\
\chi^{n+1}&\approx& \frac{1}{\omega_j-\alpha}\,\chi^n.
\end{eqnarray}      
Using the eigenvector $\chi^n$, which is proportional to $\Phi_j$, one can 
also find eigenvalue $\omega_j$ by
\begin{equation}
\omega_j=\frac{\chi^{n\dagger}H\chi^n}{\chi^{n\dagger}\chi^n}.
\end{equation}
If $\omega_j$ is close enough to $\alpha$, then one has a self consistent 
solution. This method has the benefit of directly singling out the eigenvalue 
closest to the initial guess, rather than finding the largest eigenvalue as in
the case of straight forward iteration. There is only one matrix inversion 
involved. Distribution of the grid points in momentum space is done by the 
arctangent mapping, as in the solution of the Dyson-Schwinger equation. The 
typical number of momentum space grid points used in order to obtain stable 
solutions is around 40. For angular integrals, 20 grid points which are 
linearly distributed in interval $(0,\pi)$ proved satisfactory.
\subsection{Confinement test}
Since the test of confinement involves a highly oscillatory integral 
Eq.~\ref{Delta}, we have used a large number, 30,000, of linearly distributed 
grid points. The upper limit of the momentum space integral, which is highly 
convergent, is $400\sigma$. These choices allow one to calculate the fourier 
transform(Eq.~\ref{Delta}) confidently within the timeframe shown in 
Figures~\ref{moft1.plt},\ref{moft1.5.plt}, and \ref{moft1.97.plt}.

\end{document}